\documentclass[letterpaper]{article} 
\usepackage{amsmath,amssymb}
\usepackage{aaai2026}
\usepackage{times}  
\usepackage{helvet}  
\usepackage{courier}  
\usepackage[hyphens]{url}  
\usepackage{makecell}  
\usepackage{graphicx} 
\usepackage{natbib}  
\usepackage{caption} 
\usepackage{subcaption}
\usepackage[skip=5pt]{caption}
\usepackage{float}
\usepackage{amsmath}

\usepackage{algorithm}
\usepackage{algorithmic}
\usepackage{newfloat}
\usepackage{listings}
\usepackage{multirow}
\usepackage{graphicx}
\usepackage{subfig} 
\DeclareCaptionStyle{ruled}{labelfont=normalfont,labelsep=colon,strut=off} 
\floatstyle{ruled}
\newfloat{listing}{tb}{lst}{}
\floatname{listing}{Listing}
\usepackage{xcolor}

\title{The Times They Are A-Changin': Tracking Global Trends \\ through the Lens of Web Search}
\author{
    Anju Bhuiya\textsuperscript{\rm 1},
    Hasim Ali\textsuperscript{\rm 2},
    Kovvada Dhyana Vardhan\textsuperscript{\rm 1},
    Abhijnan Chakraborty\textsuperscript{\rm 1}
}
\affiliations {
    \textsuperscript{\rm 1}IIT Kharagpur,
    \textsuperscript{\rm 2}NIT Durgapur
}

\usepackage{bibentry}
\nocopyright
\begin{document}

\maketitle
\begin{abstract}
Human societies have always left behind traces of their collective priorities in the forms of cave drawings and stone carvings to chronicles on papyrus and printed books. In the digital era, these traces increasingly emerge through social media footprints and web search behavior at the global scale.
We construct the first large-scale longitudinal dataset of Google {\it ``Year in Search''} trends, with an average of 52 countries represented per year and 15 years (2011--2025) to characterize how collective public attention evolves across countries and time.
Because search engines like Google and Bing do not publicly release raw search logs, the {\it ``Year in Search"} provides a rare cross-national archive of annually salient search interests. 
The dataset is supplemented with a manually curated, unified ten-category semantic taxonomy with strong inter-annotator agreement.

We focus our analyses on four of these categories
-- Information Seeking, Personality, Entertainment, and Sports -- to investigate how collective attention varies across geographic and socio-cultural contexts.
Our analyses reveal a consistently {\it glocalized structure of collective attention}: countries converge strongly in profession-level personality attention while remaining differentiated in specific individuals they search for. However, profession-level similarity is weakly associated with geographic region and income group. Information-seeking behavior is globally dominated by politics and health, while regional distributions show varying degrees of divergence. Crisis periods are associated with both shared and regionally differentiated shifts. 
Entertainment and sports search attention exhibit a glocalized pattern: globally prominent interests coexist with regional differentiation.

Overall, our findings suggest that web search data provide a valuable computational lens for observing collective attention at scale, acting as a living archive of how societal priorities and interests vary across countries and change over time.
\end{abstract}
\renewcommand{\thefootnote}{}
\footnotetext{\textcolor{red}{\footnotesize
This paper has been accepted at the 21st International AAAI Conference on Web and Social Media (ICWSM 2027). 
}}
\renewcommand{\thefootnote}{\arabic{footnote}}
\section{Introduction}
Human history has always left behind traces of what people collectively valued. 
From stone inscriptions and pottery shards to manuscripts and books, each era documented its priorities through the materials available to it. In the digital era, the archive of collective interest has shifted from physical debris to digital footprints. In the digital age, this documentary function has been substantially taken over by web search behavior. Through billions of daily queries, people globally contribute to a continuous, population-scale record of \textit{what they want to know}, \textit{who they find notable}, and \textit{what cultural content} captures their attention. Unlike social media, where expression is shaped by the anticipated reactions of a public audience, search queries capture unprompted, organic intent -- what people actually seek rather than what they choose to display.  This distinction makes web search data a uniquely valuable signal for studying the evolution of collective attention at scale.

Google's ``\textit{Year in Search}'' platform 
provides an annually curated summary of the most prominent and rapidly rising search interests across countries. Given Google's global scale and reach, its search trends serve as a broad proxy for public curiosity and cultural attention. Prior work has demonstrated the value of Google Trends data across domains, including economics~\cite{choi2012predicting}, finance~\cite{kristoufek2013bitcoin}, public health~\cite{wang2023real,mavragani2019google}, and population well-being during crises~\cite{brodeur2021covid}. Cebrian 
{\it et al.}~\cite{cebrian2023google} further established Google Trends as a valuable resource for studying collective attention patterns. Despite this breadth of application, most prior work examines short-term patterns, isolated events, or narrow thematic domains within single countries. The long-term, cross-national evolution of collective attention across multiple cultural domains through search behavior has hitherto remained unexplored.

A fundamental obstacle to such analyses is access to data. Google (or other web search engines like Bing) does not publicly release raw search logs. Standard Google Trends APIs\footnote{\url{https://developers.google.com/search/apis/trends}} provide relative keyword popularity indices for 
user-specified queries, but these require prior knowledge of what to search for. Also, they do not capture the open-ended landscape of what populations find most salient in any given year. The ``Year in Search" platform occupies a different position: it provides Google's own annually curated summaries of the most prominent trending interests, country by country, and is 
publicly accessible. This makes it a rare reproducible and cross-national archive of search salience constructed without proprietary data access. To the best of our knowledge, no prior work has systematically collected, unified and further enhanced this resource into a large-scale longitudinal dataset.

We address this gap by constructing a longitudinal dataset from Google {\it ``Year in Search"} data.
We focus our analyses on the 2011 -- 2025 period, where country coverage is sufficiently consistent for cross-national comparison. The platform's trend labels are highly heterogeneous and multilingual; we develop a unified semantic taxonomy to enable systematic analysis across countries and years. 
Beyond topic-level categorization, we further annotate entities within thematic domains to support fine-grained analysis of professions, entertainment genres, sports interests, and ideological attention across countries and time.
Our analyses focus on Information Seeking, Personality, Entertainment, and Sports, as these domains capture complementary dimensions of public attention, ranging from informational needs and symbolic public figures to cultural preferences and regional identity.

The central question we address in this work is {\it whether digital collective attention is globally convergent, locally differentiated, or simultaneously both}. Since web search behavior emerges through the interaction between local cultural contexts and globally shared media ecosystems, web search data a natural setting for studying the relationship between cultural convergence and regional specificity. Building on Robertson's seminal works on {\it glocalization}~\cite{robertson1992globalization,robertson1995timespace}, we investigate whether collective attention exhibits globally shared or regionally differentiated structures across domains.


To study these dynamics across informational, symbolic, and cultural domains, we organize our analyses around 
the following three research questions:

\textbf{RQ1:} What information-seeking patterns dominate global search attention? How do major crises reshape these patterns across regions?

\textbf{RQ2:} How is collective personality attention structured across countries? Does it reflect globalization or glocalization?

\textbf{RQ3:} Do entertainment and sports attention patterns combine global convergence with regional specificity?

\vspace{1mm} \noindent
In summary, our work makes three principal contributions. 
\begin{itemize}
    \item We construct and publicly release the first large-scale longitudinal dataset of Google {\it ``Year in Search"} trends, supplemented with manually curated unified semantic taxonomy. Such a unified semantic taxonomy enables 
    cross-national and longitudinal analysis of heterogeneous trend topics.

    \item  Focusing on Information Seeking, Personality, Entertainment, and Sports, we examine how global and local forces jointly structure collective attention across countries and time.  

    \item  Our analyses reveal a consistently glocalized structure of digital public attention: countries converge strongly in profession-level personality attention while remaining highly differentiated in specific named individuals, a dissociation that persists across geographic and economic boundaries. Information-seeking behavior is globally dominated by politics and health, with major crises producing substantial yet regionally differentiated shifts in public attention. Entertainment preferences converge around globally shared genres while preserving regional variation in secondary cultural preferences, and sports attention centers on football globally while sustaining strong continent-specific signatures. Together, these findings suggest that web search behavior provides a valuable computational lens into evolving societal priorities, cultural identity, and collective curiosity at global scale.
\end{itemize}

To support reproducible research on collective attention and cultural dynamics, the collected Google {\it ``Year in Search"} dataset is available on request. 
We hope this resource facilitates future computational social science research on large-scale digital behavior, cultural evolution, and global public attention.
\section{Related Work}
Google Trends data has been widely used to study aggregate behavioral and attention patterns across domains including economics~\cite{choi2012predicting}, finance~\cite{kristoufek2013bitcoin}, conservation~\cite{nghiem2016google}, and public health~\cite{ginsberg2009detecting,wang2023realtime,brodeur2021covid}. Prior work demonstrates that search behavior captures meaningful responses to economic shifts, crises, and public events. Cebrián and Doménech~\cite{cebrian2023google} further evaluated Google Trends as a longitudinal social science data source for studying collective attention. However, most existing studies rely on researcher-specified keyword queries and focus on relatively short time windows or single-country settings. Our work differs by analyzing Google’s annually curated {\it ``Year in Search"} summaries to study long-term cross-national patterns of collective attention across multiple cultural domains.

More broadly, computational social science research increasingly uses large-scale digital traces to study societal behavior, cultural dynamics, and collective attention~\cite{lazer2009computational}. Prior work has shown that collective attention in digital environments follows regular temporal dynamics~\cite{wu2007novelty}. Large-scale digital traces have also been used to reveal cross-cultural behavioral patterns across countries~\cite{golder2011diurnal}. Recent work has analyzed large-scale interactions with conversational AI systems and identified dominant thematic patterns in user information-seeking behavior~\cite{karnam2026bowling}. Our work differs by focusing on web search behavior, which provides a less socially mediated and substantially longer-term signal of collective curiosity across countries and cultural contexts. Compared to social media engagement, which is often shaped by audience visibility and performative expression, search queries provide a more direct signal of informational intent.

Computational studies of cultural variation increasingly use behavioral data to characterize how preferences converge or diverge across societies. The Moral Machine experiment~\cite{awad2018moral}, for example, demonstrated that human moral preferences exhibit both globally shared patterns and substantial regional variation. Our work adopts a related cross-national perspective but focuses on passively generated behavioral traces rather than experimentally elicited responses. The theoretical framework guiding our analysis is glocalization~\cite{robertson1992globalization,robertson1995timespace}, which describes the coexistence of globalizing and localizing forces in cultural systems. We operationalize this concept computationally by distinguishing between broad structural attention patterns and localized entity-level preferences across countries and domains of public attention.

Prior research has also shown that collective attention dynamically responds to major societal disruptions. Search behavior shifts substantially during public health crises and emergencies~\cite{brodeur2021covid,wang2023realtime}, making
digital trace data a valuable lens into evolving public priorities. Most existing
studies, however, analyze relatively short time periods due to limitations in
longitudinal behavioral data availability. By constructing a longitudinal
cross-national dataset from Google {\it ``Year in Search"}, our work enables the study of long-term structural regularities in collective attention across multiple thematic domains, including information-seeking, personality attention,
entertainment, and sports.

\section{Dataset Construction}
\begin{figure}[t]
  \centering
  \includegraphics[width=0.9\columnwidth]{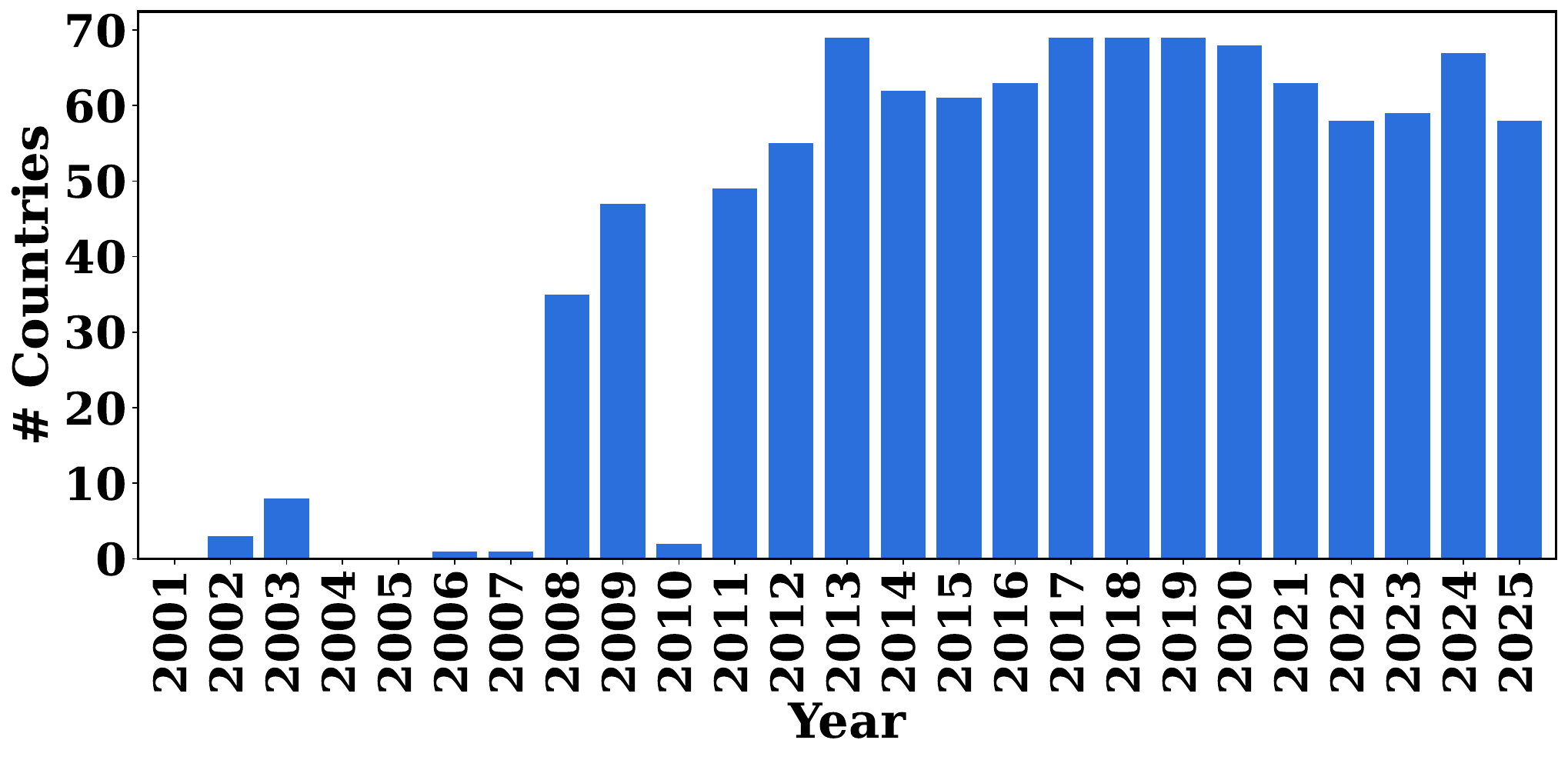}
  \caption{Number of countries with available \textit{Year in Search} data per year, 2001--2025. Coverage is sparse before 2008 and stabilizes from 2011 onward.
    }
  \label{fig:locations_per_year}
\end{figure}

To analyze global public attention over time, we constructed a longitudinal dataset from Google’s {\it ``Year in Search"} platform spanning 2001--2025~\cite{googletrends}. Unlike raw Google search logs, which are not publicly accessible, the platform provides annually curated summaries of the most prominent and rapidly rising search interests across countries. We systematically collected and unified these publicly released yearly trend summaries into a cross-national longitudinal dataset for computational analysis.

The website distills billions of searches into yearly lists of trending search topics across countries. Each topic lists up to ten prominent trending terms. 
We collected all available country-specific pages in January 2025 using Selenium-based web automation and recorded the topics exactly as presented on the website. We excluded aggregated global-view entries because they are not attributable to individual countries and would distort cross-national comparisons. Since some country pages are published in local languages, we translated all non-English topics and associated search terms into English using the Google Translate API to enable systematic comparison across countries.

Figure~\ref{fig:locations_per_year} illustrates the growth of country coverage over time. Fewer than ten countries appear before 2008, increasing to 35 in 2008 and stabilizing at 55--70 from 2011 onward. Because pre-2011 coverage is too sparse for reliable cross-national inference, our primary analyses focus on the 2011--2025 period. To further ensure temporal consistency, we restrict analysis to countries with at least 12 years of data within this window. The final dataset comprises trend lists averaging 52 countries per year, yielding 6{,}772 trending topic entries in total.

The {\it ``Year in Search"} platform reflects Google’s curated summaries of salient yearly searches rather than the complete distribution of search activity. Consequently, our analyses characterize relative patterns of collective attention and culturally salient interests rather than absolute search volumes or comprehensive user-level behavior. 
\section{Topic Taxonomy}\label{topic}
The website publishes topics in heterogeneous and country-specific forms, ranging from generic query patterns (``How to'', ``What is'') to highly localized categories (``Canadian News'', ``Dutch YouTubers''). Such heterogeneity makes direct cross-national and longitudinal comparison challenging. To enable systematic comparative analysis across countries and years, we organized the published topic labels into a unified semantic taxonomy.

For this classification, we drew on the query-intent framework~\cite{pound2010ad}. The framework distinguishes among entity, type, attribute, relation, and keyword queries based on the underlying retrieval target. Building on this intent perspective, we defined ten high-level semantic categories capturing the primary thematic intent of the published trend topics: \textit{Information Seeking}, \textit{Personality}, \textit{Entertainment}, \textit{Aggregate}, \textit{News \& Events}, \textit{Sports}, 
\textit{Food \& Cooking}, \textit{Technology \& Brands}, \textit{Travel \& Places}, and \textit{Health}. 

We distinguish \textit{Information Seeking} from the remaining ten categories based on whether the published topic specifies a target of attention. Information Seeking comprises generic interrogative topics such as ``How to'', ``What is'', ``Who'', etc., where the retrieval target is a fact, explanation, or other information. The remaining categories are entity-anchored, as the published topic specifies the object or subject of interest. For example, ``Dutch YouTubers'', ``World personalities'' specify a class of individuals mapped to the \textit{Personality} category, while ``TV Shows'', ``Song Lyrics'' specify entertainment-related objects. \textit{Aggregate} contains published summary topics, such as ``Overall''.

\begin{table}[t]
\centering
\begin{tabular}{ll}
\hline
\textbf{Topic} & \textbf{Label} \\
\hline
How to & Information Seeking \\
Dutch YouTubers & Personality \\
Canadian News & News \& Events \\
Trends of the year & Aggregate \\
Events of the year & News \& Events \\
Who & Information Seeking \\
World personalities & Personality \\
Recipe & Food \& Cooking \\
TV series and programs & Entertainment \\
Phones & Technology \& Brands \\
\hline
\end{tabular}
\caption{Examples of mapping Google Trends topics to thematic categories.}
\label{tab:topic_labels}
\end{table}

\begin{figure}[t]
  \centering
  \includegraphics[width=\columnwidth,keepaspectratio=true]{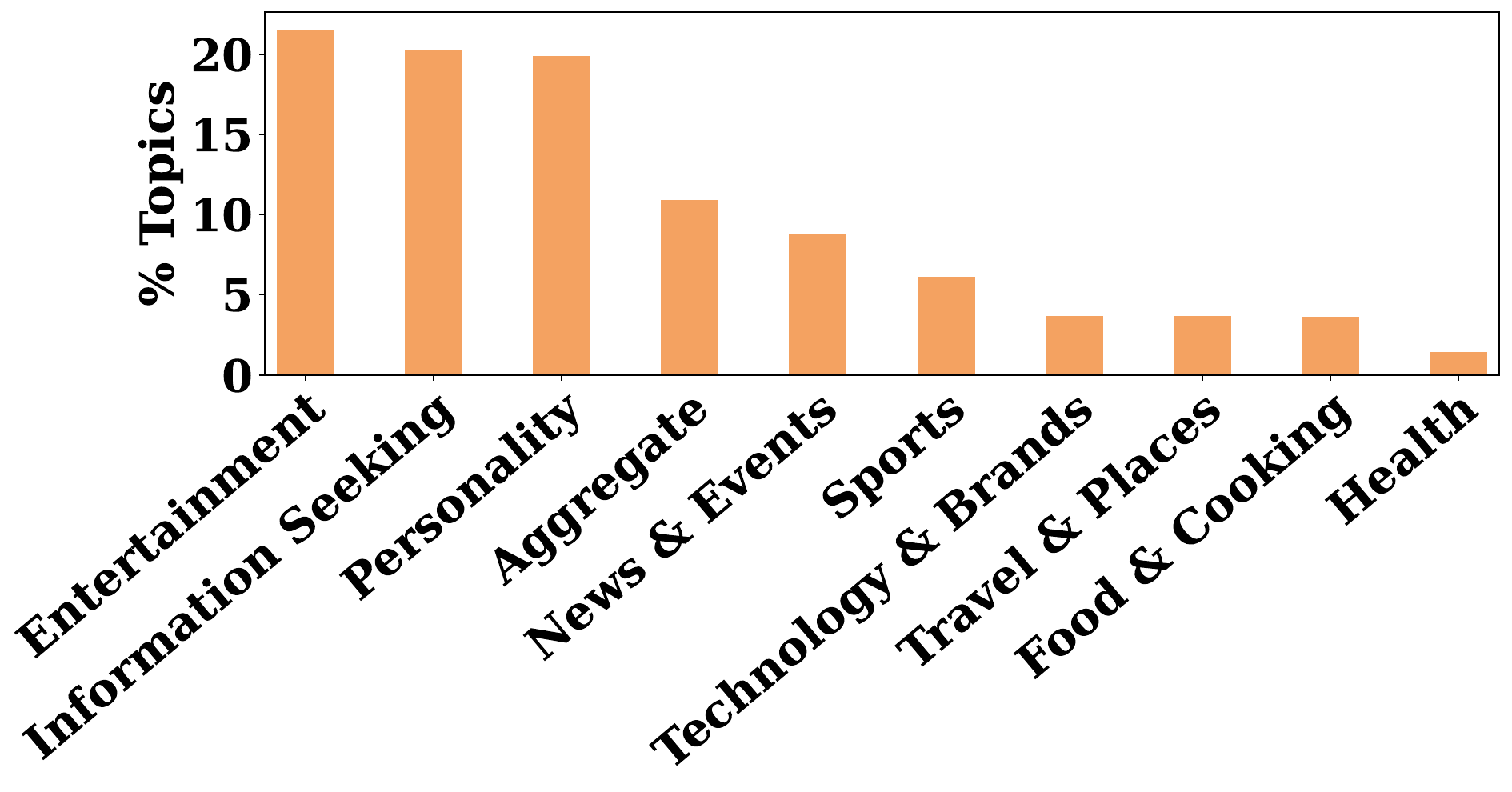}
  \caption{Distribution of Google Year in Search topics across the thematic taxonomy. 
    }
  \label{fig:bucket_distribution}
\end{figure}

Table~\ref{tab:topic_labels} presents representative examples of the mapping from original Google topic labels to the unified thematic taxonomy. 
Three annotators independently annotated the topic labels according to the predefined taxonomy. 
Pairwise Cohen's $\kappa$ scores were 0.847 (Annotators 1--2), 0.834 (Annotators 2--3), and 0.821 (Annotators 1--3), with a mean agreement of $\kappa = 0.834$, indicating substantial inter-annotator agreement. Remaining disagreements were resolved through discussion until a consensus was reached.

The identified categories are shown in Figure~\ref{fig:bucket_distribution}. The most prominent category is \textit{Entertainment} (21.56\%), \textit{Information Seeking} (20.27\%) and \textit{Personality} (19.89\%).
We focus our subsequent analyses on Information Seeking, Personality, Entertainment, and Sports because these domains directly reflect human curiosity, cultural preferences, and collective attention across countries and years.


These thematic categories provide the semantic foundation for the analyses that follow. For the regional analyses, countries were grouped into the six continents.
Africa, Asia, Europe, North America, South America, and Oceania. 
We first examine information-seeking behavior to understand how public concerns and attention patterns evolve across regions and over time. We then analyze collective personality attention to investigate cross-national patterns of attention toward public figures and symbolic domains. Finally, we study entertainment and sports attention as culturally grounded domains that reflect regional preferences, media ecosystems, and collective identity in digital search behavior.
 \begin{figure*}
    \centering
    \includegraphics[width=1\linewidth,height=0.22\textheight,keepaspectratio]{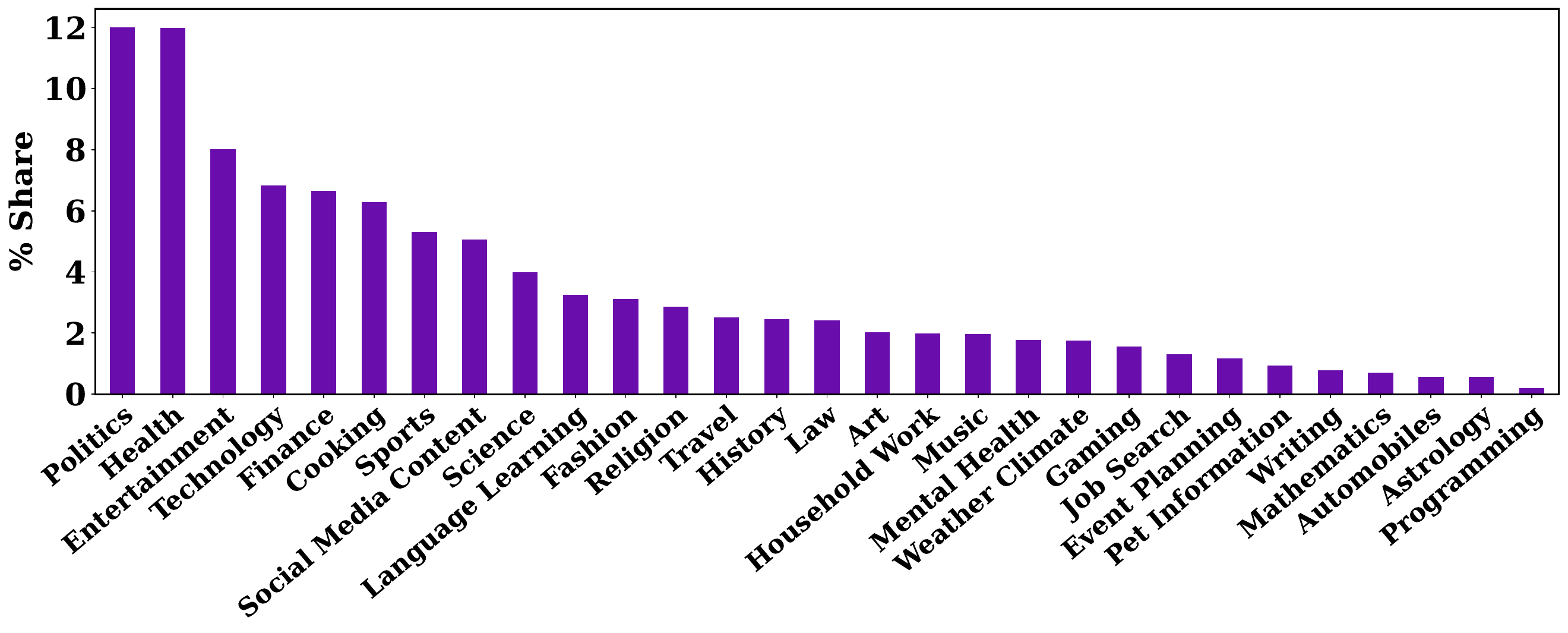}
    \caption{Distribution of categories of Information seeking queries in Google Trends Dataset.}
    \label{fig:query_dist}
\end{figure*}
\section{A World of Questions}
Search queries offer a window into
everyday concerns, interests, and public attention across countries and time. 

This section focuses on the \textit{Information Seeking} category defined above. To analyze what people ask when they pose open-ended questions, we annotated individual queries nested beneath these topics.
Three annotators manually assigned each query to one of 29 semantic categories following recent large-scale thematic labeling frameworks used for conversational AI interaction analysis~\cite{karnam2026bowling}.
The categories capture the dominant semantic intent of each query. Three annotators independently annotated the queries, with disagreements between the first two annotators resolved by the third author. Pairwise Cohen's $\kappa$ was computed before conflict resolution, with an overall $\kappa$ of $0.71$, indicating substantial agreement.

\subsection{Global Information-Seeking Patterns}

Figure~\ref{fig:query_dist} shows the global distribution of the 29 semantic categories assigned to Information Seeking queries.
\textit{Politics} ($12.01\%$) and \textit{health} ($11.99\%$) emerge as the dominant forms of global information-seeking. This resonates with recent evidence from conversational AI interaction logs, in which health likewise emerged as the most frequently queried topic~\cite{karnam2026bowling}. Yet the symmetry with politics in our search-trends data adds a layer that is absent from platform-specific studies. This suggests that politics constitutes a particularly prominent component of large-scale public search attention.

\textit{Entertainment} (8.01\%), \textit{technology} (6.82\%), \textit{finance} (6.66\%), and \textit{cooking} (6.29\%) 
also constitute substantial portions. This reflects the coexistence of practical, economic, and recreational information needs in everyday 
online behavior. Sports and social media content further highlight the increasing role of participatory digital culture in shaping public curiosity. 

Beyond these dominant categories, the long-tail distribution reveals the breadth of human information needs, spanning science, religion, travel, law, household work, gaming, programming, mental health, and creative activities. 
Notably, \textit{mental health} queries represent a relatively small proportion ($1.78\%$) of search activity compared to observations from analyzing conversational AI interaction logs~\cite{karnam2026bowling}. This difference may reflect variations in how users engage with conversational systems versus public web search environments.

These patterns provide a large-scale view of how people across the world use search engines not only to retrieve information, but also to navigate uncertainty, solve problems, pursue interests, and participate in social and cultural life.

\subsection{Regional Information Seeking Pattern}
We examine regional variation in information-seeking preferences relative to the global distribution using Kullback--Leibler (KL) divergence:
\begin{equation*}
D_{KL}(P_c \parallel P_G)=\sum_l P(l \mid c)\log
\left(\frac{P(l \mid c)}{P(l)}\right),
\end{equation*}
where $P(l \mid c)$ denotes the proportion of searched queries category $l$ within region $c$, and $P(l)$ denotes the corresponding proportion in the global distribution.
A KL divergence close to zero indicates that a continent's information-seeking composition closely resembles the global pattern, whereas higher values indicate greater regional distinctiveness.

\begin{figure}[t]
  \centering
\includegraphics[width=\columnwidth,height=0.22\textheight,keepaspectratio]{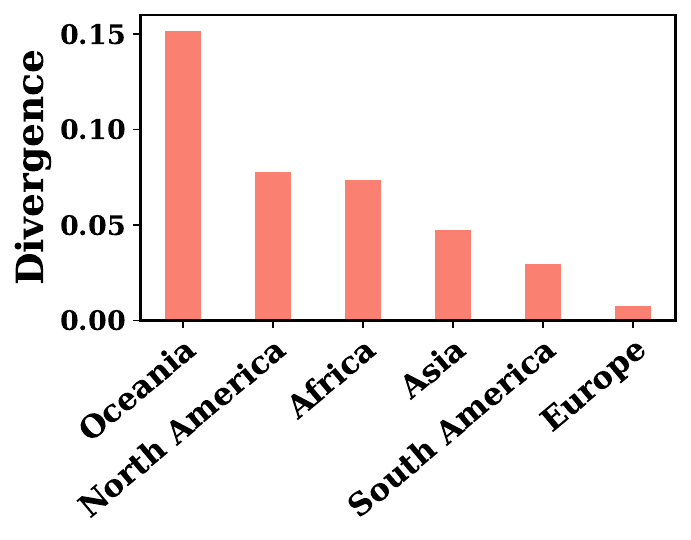}
  \caption{Divergence between regional and global information-seeking categories.}
  \label{fig:kl_div}
\end{figure}

Figure~\ref{fig:kl_div} presents the divergence between each regional distribution and the global distribution. Oceania exhibits the largest divergence ($0.152$), followed by North America ($0.078$) and Africa ($0.073$). Asia ($0.047$) and South America ($0.029$) show more moderate differences, while Europe most closely resembles the global pattern ($0.008$). These findings indicate regional variation in information-seeking preferences.

To characterize the difference observed in Oceania, we examine category-level deviations from the global distribution.
Oceania shows noticeable higher share for \textit{Social Media Content} ($+7.1\%$), \textit{Technology} ($+3.78\%$), \textit{Gaming} ($+3.24\%$) categories, accompanied by lower share for \textit{Health} ($-5.18\%$) and \textit{Politics} ($-4.18\%$) categories. This indicates Oceania's search preference skewed toward entertainment- and technology-driven curiosity rather than the health- and politics-centric pattern seen globally.

The regional analysis captures broad geographic differences; in contrast, whereas the country-level analysis reveals more localized divergence from the global pattern. At the country level, South Korea exhibits the highest KL divergence ($1.183$), followed by the United Arab Emirates ($0.735$), Pakistan ($0.698$), Saudi Arabia ($0.673$), and Hong Kong ($0.639$). South Korea's search distribution is characterized by higher shares of \textit{Language Learning} ($+26.7\%$), \textit{Technology} ($+13.2\%$), and \textit{Finance} ($+8.3\%$), among other categories. This indicates that South Korea's information-seeking distribution differs most strongly from the global pattern among the countries examined.

\color{black}

\subsection{Crisis-Driven Attention Shifts}
\begin{table*}
\centering
\resizebox{\textwidth}{!}{%
\begin{tabular}{lllllll}
\hline
\multicolumn{7}{c}{\textbf{COVID-19}} \\
\hline
\textbf{Global} & \textbf{Africa} & \textbf{Asia} &
\textbf{Europe} & \textbf{North America} & \textbf{Oceania} &
\textbf{South America} \\
\hline

Health (+12.7) &
Cooking (+8.2) &
Technology (+9.1) &
Health (+16.3) &
Health (+13.3) &
Cooking (+26.7) &
Finance (+10.9) \\

Technology (+2.6) &
Entertainment (+5.3) &
Cooking (+6.5) &
Technology (+1.9) &
Fashion (+9.8) &
Health (+18.3) &
Health (+9.1) \\

Finance (+2.3) &
Health (+4.8) &
Health (+6.0) &
Sports (+1.6) &
Language Learning (+3.2) &
Technology (+16.7) &
Mental Health (+1.4) \\

Fashion (+2.0) &
Job Search (+4.3) &
Household Work (+2.3) &
Household Work (+1.3) &
Cooking (+2.8) &
Sports (+6.7) &
Art (+0.9) \\

Household Work (+1.5) &
Technology (+3.7) &
Science (+1.8) &
Cooking (+0.7) &
Household Work (+2.8) &
Art (+3.3) &
Household Work (+0.9) \\

\hline
\end{tabular}%
}
\caption{Top rising information-seeking categories during COVID-19. Values indicate percentage-point increases in category share relative to the pre-pandemic.}
\label{tab:covid}
\end{table*}

\begin{table*}
\centering
\resizebox{\textwidth}{!}{%
\begin{tabular}{lllllll}
\hline
\multicolumn{7}{c}{\textbf{Ukraine Conflict}} \\
\hline
\textbf{Global} & \textbf{Africa} & \textbf{Asia} &
\textbf{Europe} & \textbf{North America} & \textbf{Oceania} &
\textbf{South America} \\
\hline

Politics (+6.3) &
Religion (+6.9) &
Language Learning (+6.9) &
Politics (+8.3) &
Entertainment (+6.0) &
Language Learning (+22.2) &
Music (+4.2) \\

Travel (+2.0) &
Event Planning (+4.3) &
Fashion (+5.8) &
Finance (+3.5) &
Sports (+5.8) &
Politics (+19.4) &
Fashion (+2.6) \\

History (+1.7) &
Language Learning (+4.3) &
Travel (+4.3) &
History (+3.2) &
Politics (+4.3) &
Social Media Content (+8.3) &
Politics (+2.5) \\

Religion (+1.4) &
Social Media Content (+4.3) &
Politics (+3.6) &
Fashion (+2.2) &
Travel (+2.8) &
Entertainment (+5.6) &
Astrology (+2.4) \\

Astrology (+1.1) &
History (+3.4) &
Social Media Content (+3.5) &
Science (+1.7) &
Cooking (+2.4) &
Science (+5.6) &
Science (+2.4) \\

\hline
\end{tabular}%
}
\caption{Top rising information-seeking categories during the Ukraine conflict. Values indicate percentage-point increases in category share relative to the pre-conflict period.}
\label{tab:ukraine}
\end{table*}

We further examine how major global crises reshape information-seeking priorities across regions through two case studies: the COVID-19 pandemic (2020) and the war in Ukraine (2022). For each event, we compare search-category distributions during the crisis with those in the corresponding pre-crisis period. Table~\ref{tab:covid} reports the dominant positive shifts during COVID-19, and Table~\ref{tab:ukraine} reports the corresponding shifts during the Ukraine conflict.

COVID-19 generated a broad increase in health-related searches, making \textit{Health} the dominant global shift. This pattern is consistent with the immediate information needs arising from a public health emergency. However, the regional results indicate that crisis-related attention also extended to practical and everyday activities. \textit{Cooking} appeared among the top-rising categories in every region except Africa, while \textit{Household Work} appeared in the global as well as Asia, Europe, North America, Oceania, and South America regions.  \textit{Technology} also appeared across several regions, suggesting increased attention to tools and activities associated with everyday adaptation during the pandemic. 
At the same time, the relative emphasis differed by region: Asia showed stronger increases in \textit{Technology} and \textit{Cooking}; Europe and North America showed particularly strong increases in \textit{Health}; South America displayed larger increases in \textit{Finance} and \textit{Health}; and Oceania showed comparatively large increases in \textit{Cooking}, \textit{Health}, and \textit{Technology}. 
Thus, COVID-19 was associated with both a widely shared health-related response and broader changes in practical, domestic, and technology-related information seeking.

The Ukraine conflict produced a more differentiated pattern of attention. \textit{Politics} was the dominant global increase, consistent with the conflict's direct relevance to geopolitical developments and public affairs.
However, the rising categories also extended beyond politics. \textit{History} appeared in the global, African, and European results, while \textit{Science} appeared in Europe, North America, Oceania, and South America. Europe showed particularly strong increases in \textit{Politics}, \textit{Finance}, and \textit{History}, which may reflect the region's geographic and political proximity to the conflict. Other regions displayed different combinations of rising categories. Asia showed stronger increases in \textit{Language Learning}, \textit{Fashion}, and \textit{Travel}; North America showed increases in \textit{Entertainment} and \textit{Sports}; South America showed a notable increase in \textit{Music}; and Africa exhibited a stronger increase in \textit{Religion}. These results suggest that the Ukraine conflict was associated with a combination of shared attention to public affairs and regionally differentiated increases in cultural, practical, historical, and social information seeking.

Taken together, the two case studies show that crisis-related changes in information seeking combine shared and regionally differentiated responses. COVID-19 was associated not only with increased attention to \textit{Health}, but also with rising interest in \textit{Cooking}, \textit{Household Work}, \textit{Technology}, and \textit{Finance}, reflecting a broader reorganization of everyday information needs. The Ukraine conflict showed a more geographically differentiated pattern, with \textit{Politics} as the dominant global increase and additional rises in categories such as \textit{History}, \textit{Science}, \textit{Language Learning}, \textit{Religion}, and \textit{Entertainment}. These differences are consistent with the possibility that crisis-related information needs are shaped by both the broad nature of an event and regional exposure, political context, economic conditions, and everyday practical concerns. Because the analysis captures changes in search behavior, the observed shifts can be interpreted as associations with crisis periods.

\color{black}
\begin{figure}
    \centering
    \includegraphics[width=1\linewidth]{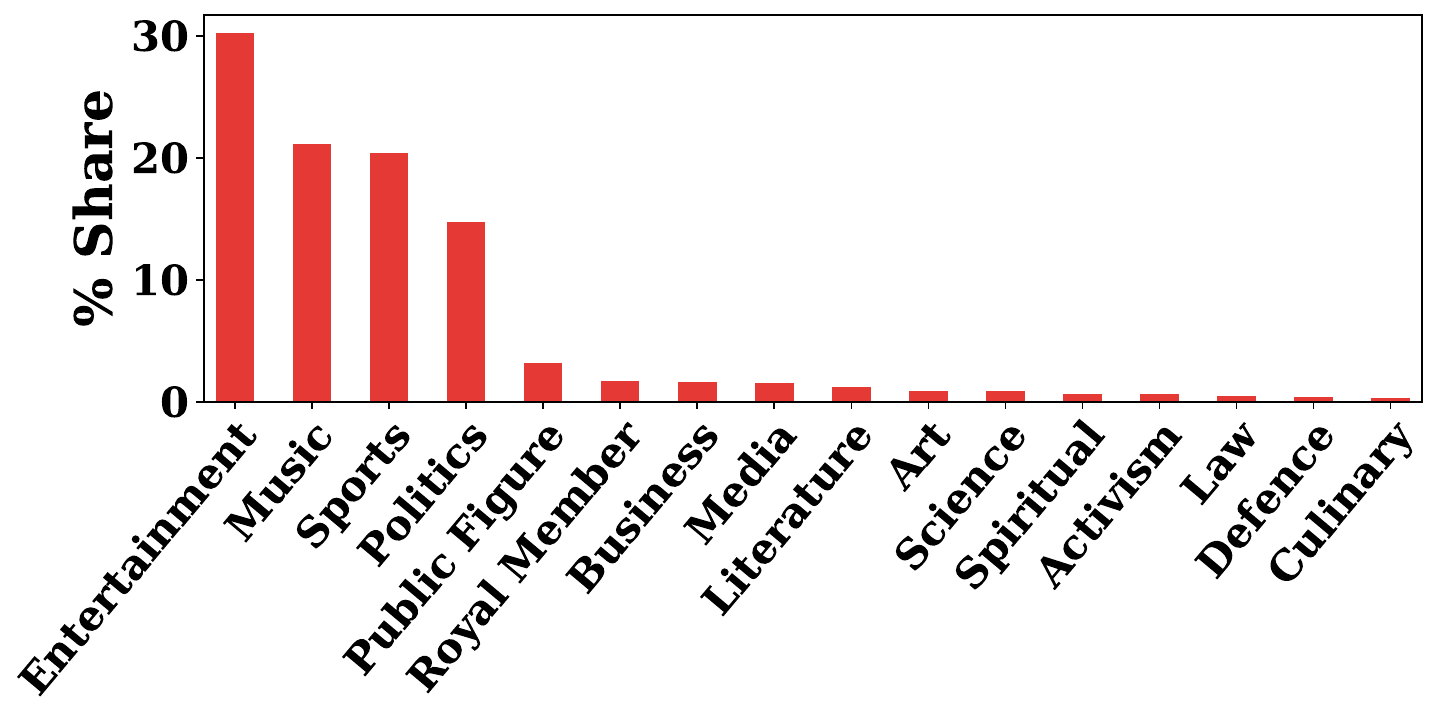}
    \caption{Distribution of professions of searched personality's professions in Google Year-in-Search data. Entertainment, Music, Sports, and Politics dominate global personality attention.}
    \label{fig:profe}
\end{figure}
\section{Global Structure of Personality Attention}














\begin{figure}[t]
  \centering
\includegraphics[width=0.7\columnwidth,height=0.25\textheight,keepaspectratio=true]{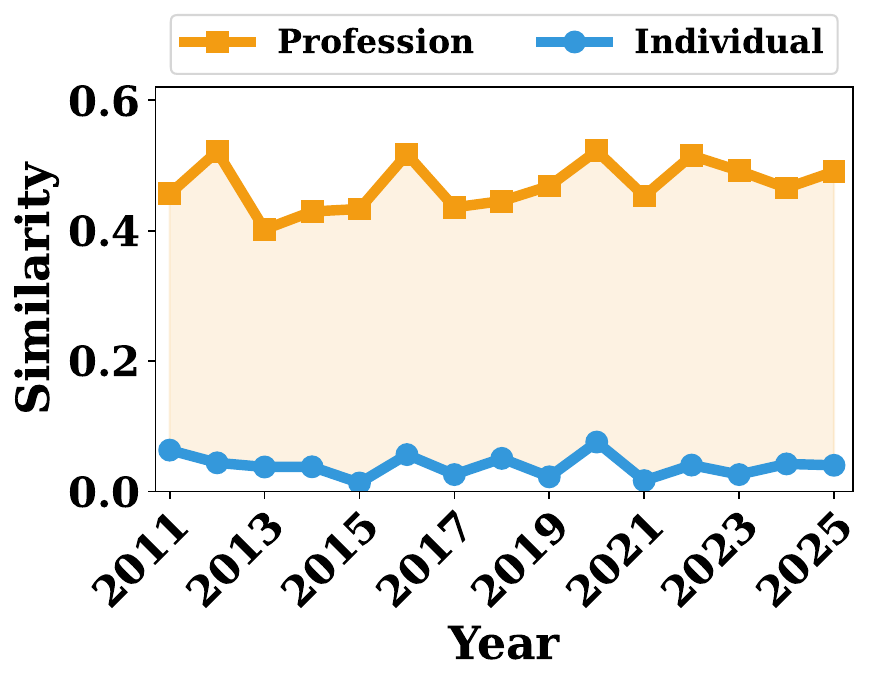}
\caption{
Temporal evolution of cross-country similarity in collective personality attention from 2011--2025. 
Profession-level similarity remains consistently high, while name-level similarity stays close to zero. This shows that countries tend to attend to similar profiles, but not necessarily the same individuals. 
}

  \label{fig:glocalization}
\end{figure}
We next examine the global structure of collective personality attention and whether countries exhibit shared symbolic attention patterns.
We analyze the profession of the personalities appearing in 
Google Year-in-Search 
trends. Each personality's profession was independently annotated by three annotators using targeted web searches, with ambiguities resolved by majority consensus. Inter-annotator reliability was substantial (Fleiss’s $\kappa$~\cite{falotico2015fleiss} = 0.79). If a searched individual does not belong to a specific industry 
but drew attention due to 
sensational events, we classified them as Public Figures.
Then, the occupation was mapped to 16 high-level profession categories, including Politics, Sports, Entertainment, Music, Business, Science, Religion, and Activism. Figure~\ref{fig:profe} presents the distribution of the searched personality's profession. 

\subsection{Glocalization of Personality Attention}
A central question in cultural globalization research is whether increased platform connectivity produces \textit{homogenization}. Robertson introduced the concept of \textit{glocalization} to explain homogenization of celebrities, where the same individuals are becoming famous everywhere 
~\cite{robertson1992globalization,robertson1995time}. Glocalization refers to the coexistence of globally shared symbolic structures with locally specific content. In our context, countries may attend to similar profiles of people, such as entertainers, athletes, politicians, and musicians, but not necessarily the same individuals. To compute this distinction, we compare two complementary Jaccard similarity measures computed annually across all country pairs.
For each year $t$ and pair of countries $(c_i, c_j)$, we define:
\[
J^{\text{prof}}_t(c_i, c_j)
=
\frac{|P_{i,t} \cap P_{j,t}|}
{|P_{i,t} \cup P_{j,t}|},
\]
where $P_{c,t}$ denotes the set of profession categories represented in country $c$'s personality entries in year $t$. Analogously, $J^{\text{name}}_t$ replaces profession sets with sets of individual person names. Both measures are bounded in $[0,1]$, where 1 indicates complete overlap, and 0 indicates no overlap.

Profession-level similarity remains consistently high across countries over time, whereas name-level similarity stays close to zero (Figure~\ref{fig:glocalization}). Profession-level similarity remains stable throughout the period, ranging from 0.41 to 0.53 (mean $=0.470$, Mann-Kendall ~\cite{mann1945nonparametric}: $\tau=+0.26$, $p=.202$).
Countries consistently attend to similar categories such as entertainers, athletes, politicians, and musicians. In contrast, name-level similarity remains near to zero throughout the period (mean $=0.040$, range $[0.013,0.076]$). Countries rarely search for the same individuals even when they exhibit similar profession-level attention structures.

The gap between the two measures averages $\Delta = 0.435$ and is highly significant (Mann-Whitney~\cite{mann1947test}: $U=225$, $p=3.4\times10^{-6}$). This dissociation provides quantitative evidence for glocalization in personality attention. The globally shared symbolic structures coexist with locally specific content.
While the glocalization analysis demonstrates strong convergence in profession-level attention structures, it does not, however, explain whether this convergence primarily follows geographic proximity or economic similarity. 
Therefore, we next examine whether these shared attention structures align with geographic proximity or economic similarity across countries.

\subsection{Cross-National Personality Attention}
\begin{figure*}[t]
  \centering
  \includegraphics[width=\textwidth,keepaspectratio=true]{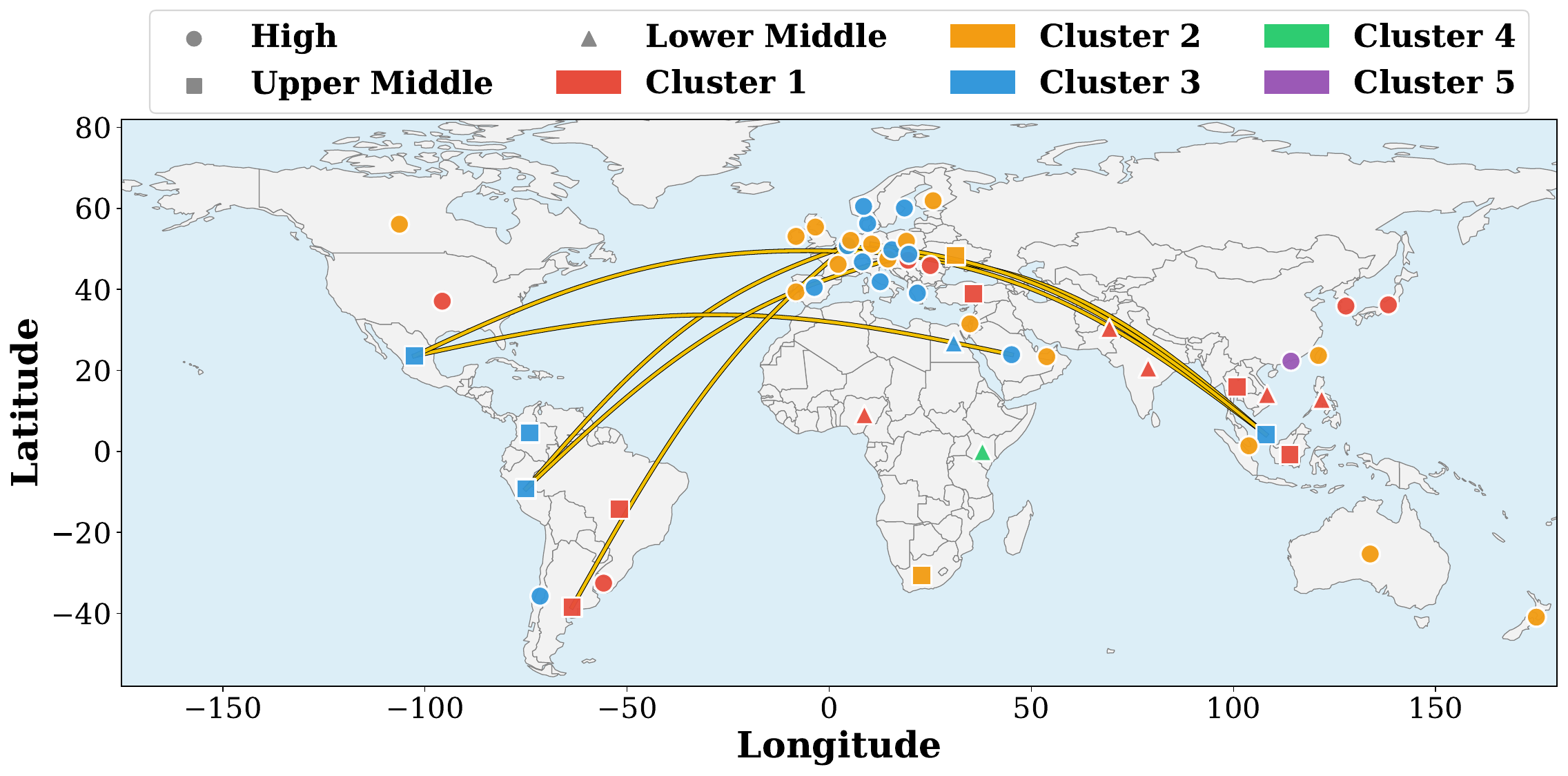}
 \caption{
Global similarity structure of profession-level collective personality attention across countries. Similar attention clusters appear across multiple continents and income groups. The profession-level attention only partially follows geographic or economic boundaries. Gold arcs highlight most similar country pairs that cross both continental and income-group boundaries.}
  \label{fig:world_person}
\end{figure*}


To analyze the profession-level attention across countries, we construct for each country a 19-dimensional profession-distribution 
vector $\mathbf{v}_c \in \mathbb{R}^{19}$. Each dimension represents the proportion of personality entries belonging to a given 
profession category aggregated across all years.

Using these profession-distribution vectors, countries are grouped through a two-stage Ward-linkage hierarchical clustering procedure on standardized features. First, countries with a $z$-scored distance from the sample centroid greater than 2.5 are identified as having unusually distinctive profession-attention profiles. This procedure identifies Kenya and Hong Kong as having distinctive profiles, which are retained as separate singleton groups. The remaining countries are then clustered, with the number of clusters selected by maximizing the silhouette coefficient ($0.0618$), subject to a minimum cluster size of three. The three core clusters are characterized by relatively higher attention to Entertainment and Music; Politics and Sports; and Music and Literature, respectively.

To examine whether geographic or economic similarity is associated with collective attention patterns, we assign countries to the regional categories defined in Section~\ref{topic} and to World Bank income groups\footnote{\url{https://blogs.worldbank.org/en/opendata/world-bank-country-classifications-by-income-level-for-2024-2025}}. 
\color{black}
Figure~\ref{fig:world_person} visualizes all countries at their geographic coordinates, with node color representing the attention cluster and marker shape representing the World Bank income group.

To quantify this relationship, we compare pairwise similarities within versus across regions and income groups using the Mann-Whitney $U$ test. Countries within the same region exhibit slightly higher similarity ($M=0.764$) than countries across regions ($M=0.753$, $p=.002$), though the effect is small ($\Delta=0.011$). 
Similarly, countries within the same income group
show greater similarity ($M=0.770$) than those across
income groups ($M=0.743$, $p<.001$), with a modest difference ($\Delta=0.028$). These results suggest that geography and income are associated with small differences in attention similarity at the profession level.

Figure~\ref{fig:world_person} additionally highlights the most similar country pairs that simultaneously cross both regional and income boundaries. Several non-trivial alignments emerge, including Belgium--Malaysia (similarity $=0.888$), Belgium--Peru ($0.875$), Malaysia--Slovakia ($0.873$), Mexico--Saudi Arabia ($0.871$), Peru--Slovakia ($0.866$), Mexico--Slovakia ($0.866$), and Germany--Malaysia ($0.865$). These pairs span multiple continents and often cross substantial income differences, with some pairs exhibiting more than a sevenfold difference in gross national income. These findings indicate that profession-level collective attention is not strongly constrained by geography or economic development. Rather, countries with substantially different political, economic, and cultural contexts can exhibit similar distributions of attention across profession categories.

\section{Entertainment Preferences}

 %

Beyond public attention to people, entertainment trends offer insight into the narratives, emotional themes, and cultural expressions that capture the collective imagination across societies.
The entertainment category includes movies, television shows, songs, and music-related topics, as represented by Google-defined topic categories.  We used movie- and television-related topics to identify movie and television content, and music-related topics, including songs and lyrics, to identify music content. The identified entities were then annotated with genre labels, and their genre distributions were examined globally and across geographic regions.

\begin{figure*}[t]
    \centering

    \begin{minipage}[t]{0.4\textwidth}
        \centering
        \includegraphics[
            width=\linewidth,
            keepaspectratio
        ]{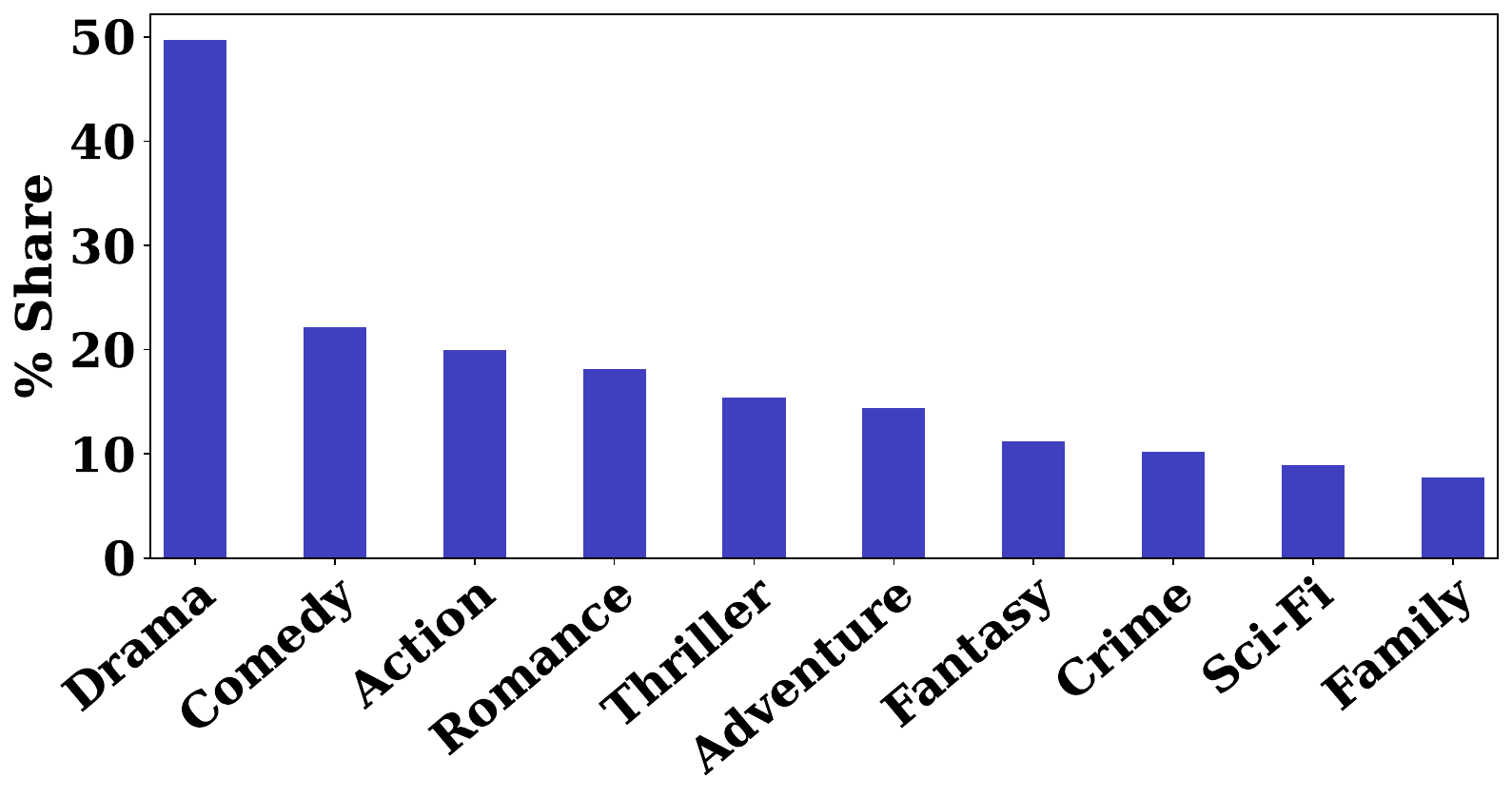}

        \caption*{(a)}
    \end{minipage}
    \hfill
    \begin{minipage}[t]{0.58\textwidth}
        \centering
        \includegraphics[
            width=\linewidth,
            keepaspectratio
        ]{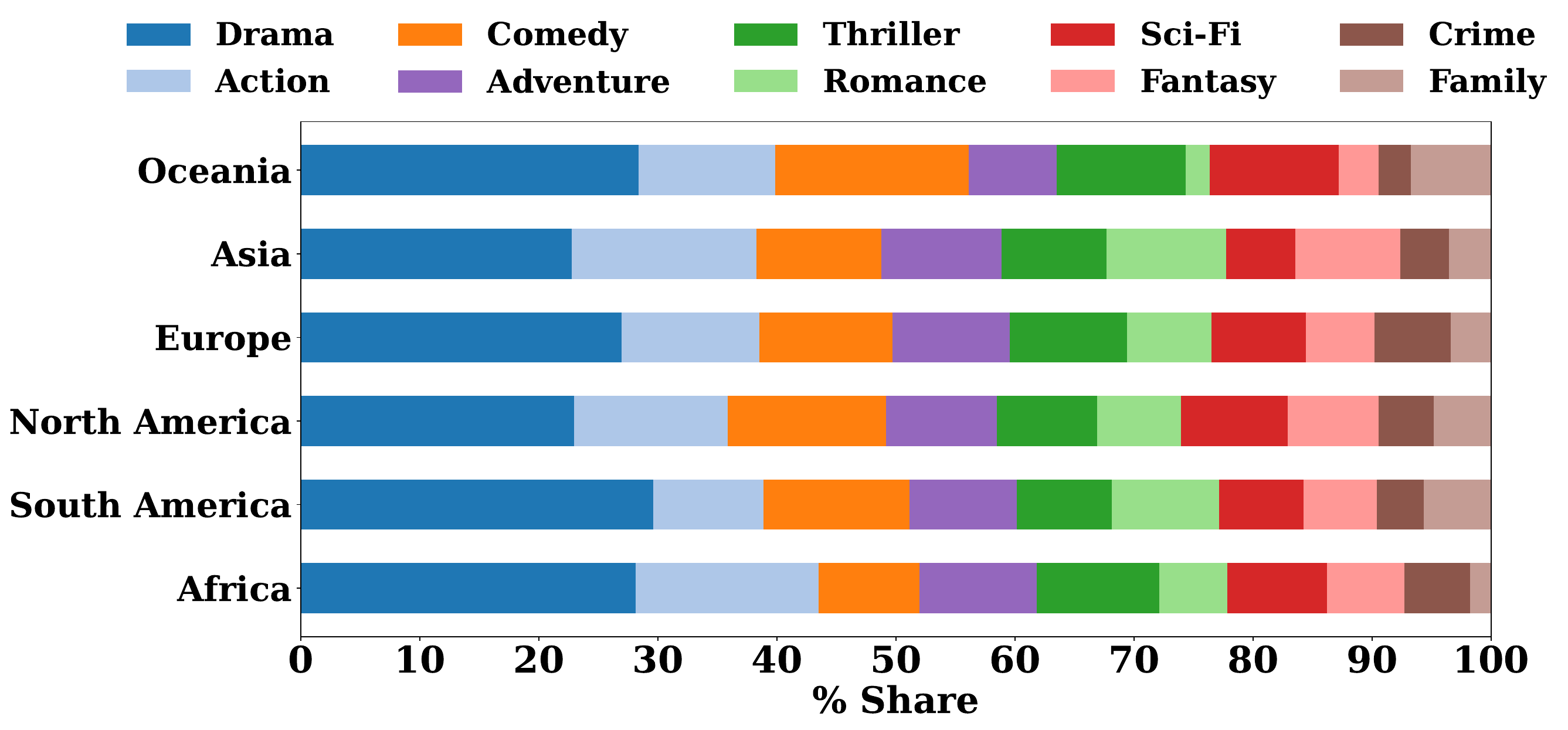}

        \caption*{(b)}
    \end{minipage}

    \caption{Global and regional distributions of genre associations among searched movies and television shows.
    (a) Drama is the most prominent genre globally, followed by Comedy, Action, and Thriller.
    (b) Drama remains prominent across regions, while the distribution of secondary genres varies geographically.}

    \label{fig:entertainment_genre}
\end{figure*}
\subsection{Movie and Television Preferences}
To analyze Movie and Television Preferences, we annotated the entities associated with movie- and television-related topics with genre labels using the {\it IMDb API}\footnote{\url{https://developer.imdb.com/documentation/api-documentation/}}.\color{black}
 Figure~\ref{fig:entertainment_genre} (a) presents the distribution of dominant entertainment genres in searched movie and television content. Drama emerges as the dominant entertainment genre, accounting for 49.7\% of all genre associations, more than twice the share of Comedy (22.2\%) and substantially exceeding Action (20.0\%). Romance, Thriller, and Adventure also occupy prominent positions, indicating strong audience preference for emotionally engaging and high-intensity narratives. In contrast, genres such as Fantasy (11.2\%), Crime (10.2\%), Sci-Fi (8.9\%), and Family (7.7\%) appear at comparatively lower levels of attention.

Figure~\ref{fig:entertainment_genre}(b) presents regional variation in entertainment genres across six geographic regions. Drama remains the leading genre across all regions, with the highest shares observed in South America (29.6\%), Oceania (28.4\%), Africa (28.1\%), and Europe (26.9\%), followed by North America (23.0\%) and Asia (22.7\%). Comedy is particularly prominent in Oceania (16.2\%), North America (13.3\%), and South America (12.2\%), while Action receives higher attention in Asia (15.5\%) and Africa (15.4\%). Adventure exhibits a relatively stable distribution across regions, ranging from 7.4\% in Oceania to 10.1\% in Asia. Romance is most prominent in Asia (10.0\%) and South America (9.1\%), whereas Thriller receives its highest shares in Oceania (10.8\%) and Africa (10.3\%). Asia shows the greatest relative attention to Fantasy (8.8\%), while Sci-Fi is more prominent in Oceania (10.8\%), North America (9.0\%), and Africa (8.4\%). Crime is comparatively more visible in Europe (6.4\%), and Family receives its highest shares in Oceania (6.8\%) and South America (5.7\%).

These findings are consistent with a glocalization pattern in entertainment attention. Although Drama provides a shared dominant genre across regions, the relative prominence of secondary genres varies geographically. Thus, globally popular entertainment appears to be structurally shared while remaining locally differentiated in its thematic emphasis.

\begin{figure*}[t]
    \centering
    \begin{minipage}[t]{0.35\textwidth}
        \centering
        \includegraphics[width=\linewidth,keepaspectratio]{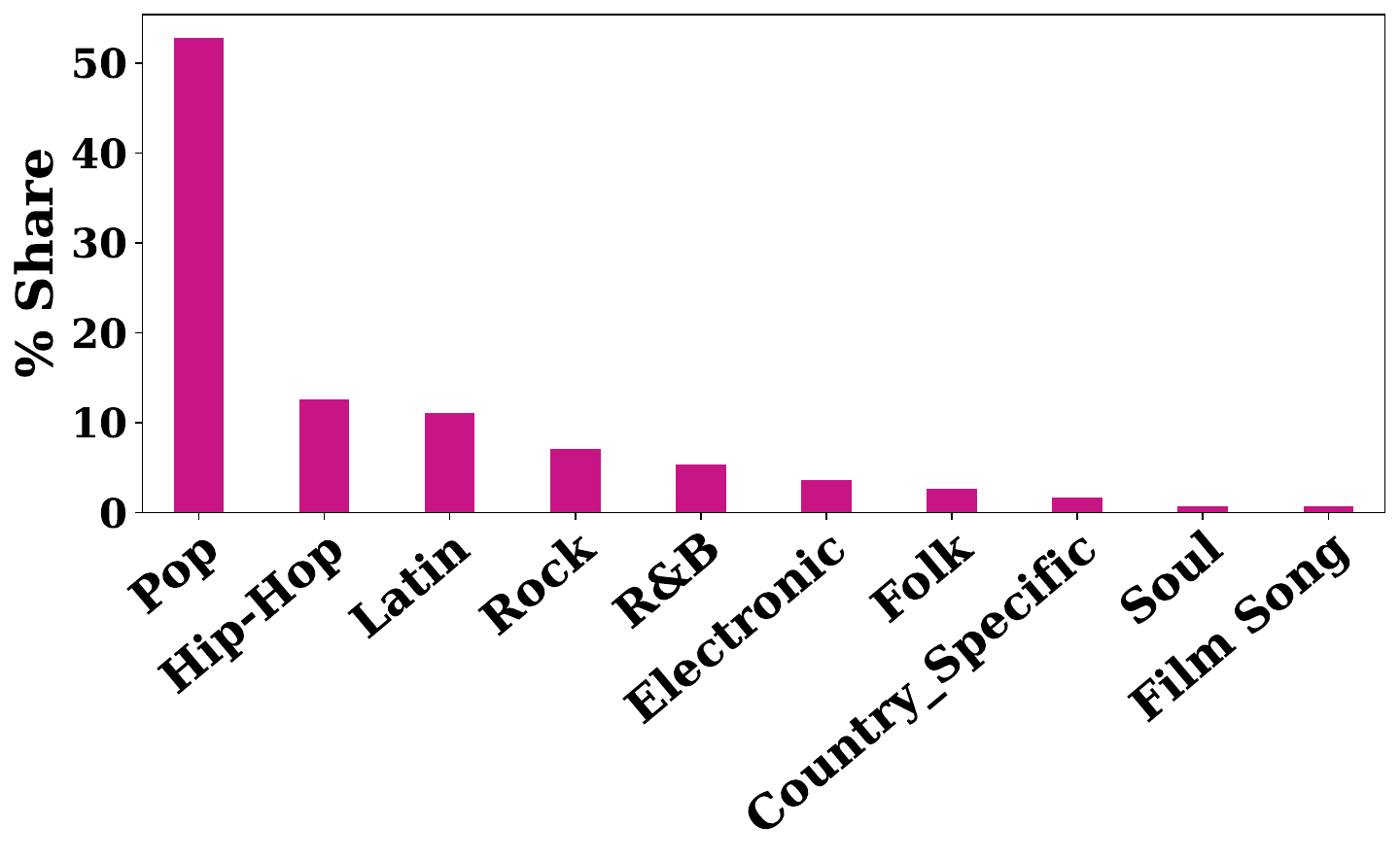}
        \caption*{(a)}
    \end{minipage}
    \hfill
    \begin{minipage}[t]{0.6\textwidth}
        \centering
        \includegraphics[width=\linewidth,keepaspectratio]{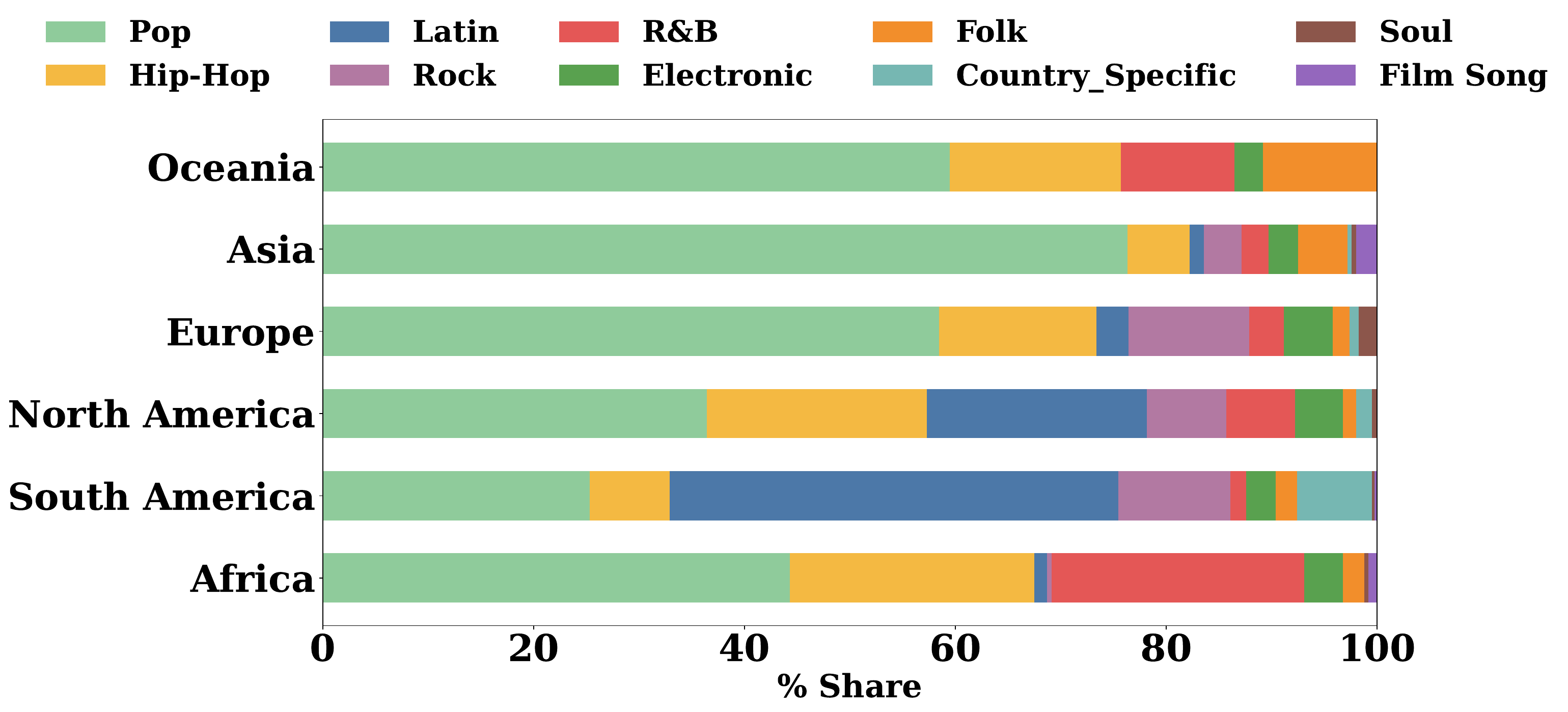}
        \caption*{(b)}
    \end{minipage}
    \caption{Global and regional patterns of music search attention represented through genre distributions. (a) Pop is the most prominent music genre globally, followed by Hip-Hop, Latin, and Rock. (b) Pop remains prominent across most regions, while the relative prominence of other genres varies by region.}
    \label{fig:music_genre}
\end{figure*}

\subsection{Music Preferences}
To analyze music preferences, we annotated the entities associated with songs and music-related topics with genre labels using Wikidata~\cite{wikidata}. Figure~\ref{fig:music_genre}(a) presents the global distribution of the ten most prominent music genres. Pop is the dominant genre, accounting for 52.76\% of all genre associations, followed by Hip-Hop (12.54\%), Latin (11.06\%), and Rock (7.03\%). R\&B represents 5.32\% of genre associations, while Electronic (3.57\%), Folk (2.64\%), and Country (1.67\%) receive comparatively lower shares. Soul (0.74\%) and Film Song (0.70\%) appear less frequently in the global distribution.

Figure~\ref{fig:music_genre} (b) presents regional variation in music genre attention across geographic regions. Pop is the leading genre in most regions, with particularly high shares in Asia (76.3\%), Europe (58.5\%), Oceania (59.5\%), and Africa (44.3\%). South America differs from this pattern, where Latin is the most prominent genre (42.5\%), followed by Pop (25.3\%) and Rock (10.6\%). North America exhibits a more diversified distribution, with Pop (36.4\%), Hip-Hop (20.9\%), and Latin (20.9\%) receiving substantial shares. Africa also shows comparatively strong attention to R\&B (24.0\%) and Hip-Hop (23.2\%), while Oceania displays notable shares for Hip-Hop (16.2\%), R\&B (10.8\%), and Folk (10.8\%). Europe shows a strong concentration in Pop, alongside Hip-Hop (14.9\%) and Rock (11.4\%). Asia is particularly concentrated around Pop, with smaller contributions from Folk (4.7\%), Rock (3.6\%), and Hip-Hop (5.9\%).

These patterns indicate that music-related attention combines a globally dominant genre structure with substantial regional differentiation. Although Pop accounts for the largest share globally and remains prominent across most regions, Latin music is particularly salient in South America, while Hip-Hop and R\&B receive greater relative attention in Africa and North America. The regional distribution, therefore, suggests the coexistence of globally shared musical attention and locally differentiated genre preferences.
\color{black}

\begin{figure*}[t]
    \centering

    \begin{minipage}[t]{0.35\textwidth}
        \centering
        \includegraphics[width=\linewidth,keepaspectratio]{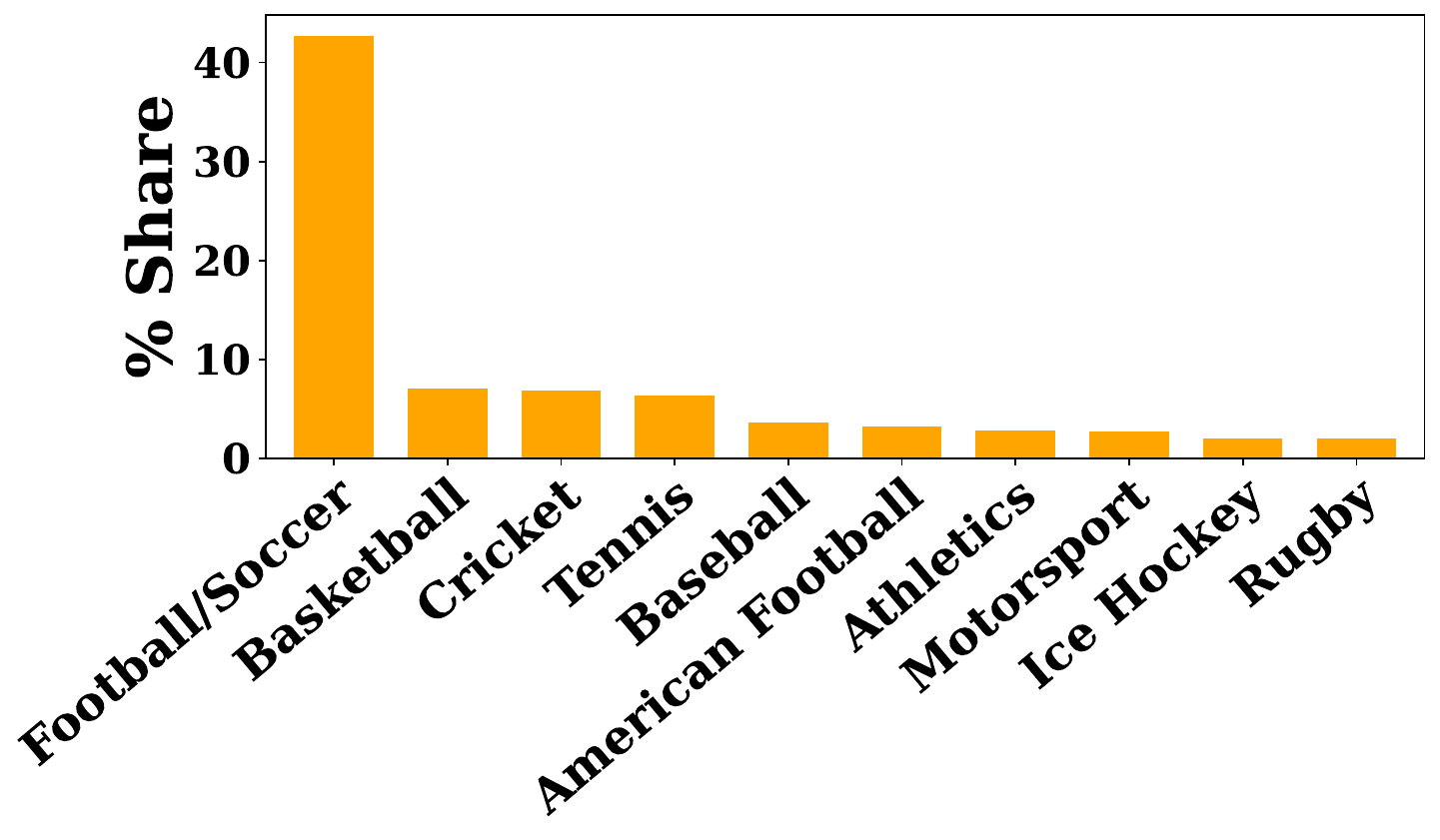}
        \caption*{(a)}
    \end{minipage}
    \hfill
    \begin{minipage}[t]{0.6\textwidth}
        \centering
        \includegraphics[width=\linewidth,keepaspectratio]{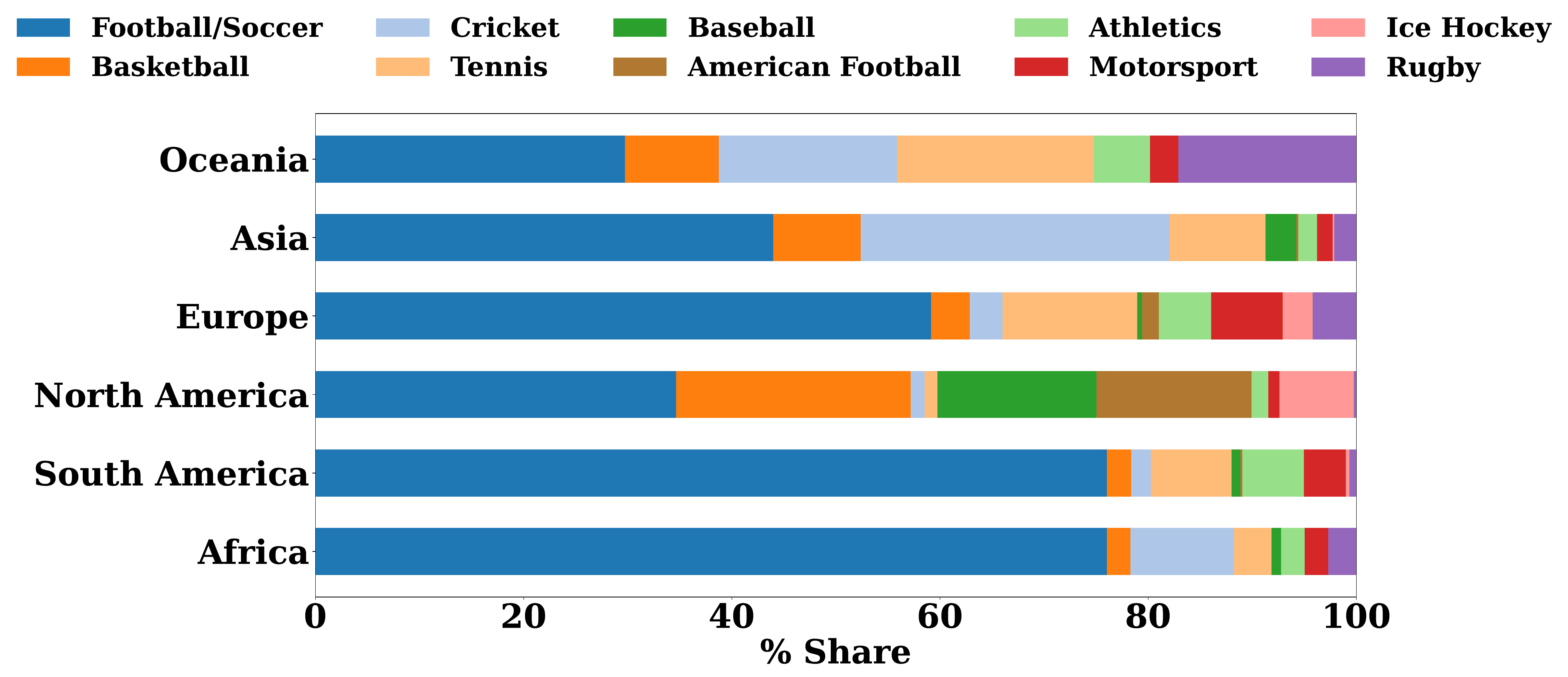}

        \caption*{(b)}
    \end{minipage}
    \caption{Global and regional patterns of sports search interest. (a) Global distribution of the ten most-searched sports, with Football/Soccer as the most prominent sport. (b) Football/Soccer remains prominent across regions, while the distribution of other sports varies geographically.}

    \label{fig:sports_combined}
\end{figure*}


\section{The World’s Sporting Pulse}
Among the thematic domains identified in our taxonomy, sports provide a lens into regionally localized forms of collective attention and cultural identity. From stadiums to online highlights, sports interests generate large-scale digital traces of shared public attention. To examine regional sporting preferences, we annotated sports-related entities in Wikidata~\cite{wikidata} and analyzed the ten most-searched sports across regions.
Figure~\ref{fig:sports_combined} (a) presents the global distribution of sports-related searches, while Figure~\ref{fig:sports_combined} (b) illustrates regional variation in sports attention.
Football/Soccer has the highest share of sports-related searches globally (42.68\%), underscoring its broad international reach. Basketball (7.10\%), Cricket (6.82\%), and Tennis (6.36\%) follow at substantially lower shares, indicating that global sports attention is strongly concentrated around Football/Soccer. At the regional level, Football/Soccer is particularly dominant in Africa (76.0\%) and South America (76.0\%), while also accounting for the largest share of attention in Europe (59.1\%) and Asia (44.0\%). Asia additionally exhibits a strong preference for cricket (29.7\%). 
North America displays a more distributed pattern, with attention concentrated across Football/Soccer (34.6\%), Basketball (22.6\%), Baseball (15.3\%), and American Football (14.9\%). Oceania likewise shows a more diverse distribution, with Football/Soccer (29.7\%), Tennis (18.9\%), Cricket (17.1\%), and Rugby (17.1\%) receiving substantial shares. Europe also exhibits comparatively greater diversification across sports, with Tennis (12.9\%) and Motorsport (6.9\%) representing notable components alongside Football/Soccer.

Together, these patterns indicate that sports-related attention combines a globally shared structure with regionally differentiated preferences. Although Football/Soccer dominates globally, the prominence of Cricket in Asia, Baseball and American Football in North America, and Tennis and Rugby in Oceania highlights regional differences in sports attention. Africa and South America show particularly concentrated attention toward Football/Soccer, whereas North America, Europe, and Oceania exhibit more diversified sports profiles. This coexistence of global convergence and regional differentiation is consistent with the broader glocalization dynamics observed throughout the paper.

\color{black}
\section{Conclusion}

Using longitudinal Google {\it ``Year in Search"} data, this work provides a large-scale cross-national analysis of how digital collective attention evolves across countries, cultural domains, and historical periods. Across information-seeking, personality attention, entertainment, and sports, we find a consistent glocalized structure of public attention: countries frequently converge in broad thematic and symbolic attention patterns while remaining strongly differentiated in specific cultural content and locally salient entities.

This convergence-divergence dynamic is most visible in personality attention, where countries share similar distributions of public figure categories despite searching for highly localized individuals. The same structural pattern extends across information-seeking behavior, entertainment preferences, and sports attention. Global crises reshape public attention worldwide, yet regional contexts mediate how those disruptions are interpreted and prioritized. Entertainment and sports similarly exhibit globally shared structures alongside persistent regional specialization.

Together, these findings suggest that digital public attention is neither fully globalized nor purely fragmented. Instead, attention patterns emerge through the interaction between globally shared media ecosystems and local cultural contexts. In this sense, glocalization is not confined to a single domain of online behavior but constitutes a broader organizing principle of digital collective attention.

Methodologically, our work demonstrates the value of publicly accessible web search archives for computational social science research. Because Google does not publicly release raw search logs, the {\it ``Year in Search"} platform provides a rare longitudinal and cross-national record of publicly salient interests. By combining this archive with a unified semantic taxonomy, entity-level annotation framework, and cross-national analysis pipeline, we provide a reproducible methodology for studying large-scale societal attention dynamics over time.

This work also carries important limitations. The {\it ``Year in Search"} platform reflects Google's curated summaries rather than the complete distribution of search activity, and the available data remains aggregated at the country level. Future work could incorporate finer temporal resolution, demographic information where available, and complementary digital trace sources to better understand how collective attention evolves across platforms and populations.

More broadly, this work highlights the potential of web search behavior as a computational lens into evolving human curiosity, cultural identity, and societal priorities. As digital traces increasingly document everyday life, longitudinal search behavior offers a unique opportunity to study how societies collectively respond to crises, cultural change, and global connectivity across time. We hope that the released dataset and semantic annotation framework facilitate future computational studies of collective attention, cultural evolution, and large-scale digital behavior across societies and historical periods.

\section{Limitations}
Our dataset is based on Google ``Year in Search'' data and is consequently limited by Google's country coverage and publication practices. Comparable records are not continuously available for all countries and years. For example, countries such as China and Russia are not included in our analyses. Consequently, our analysis is limited to the countries and years represented in Google's published records.
\color{black}

\section*{Acknowledgments}

We used large language models only for limited editorial assistance, including grammar correction, sentence refinement, and rephrasing. All experiments, analyses, annotations, interpretations, and the primary manuscript preparation were conducted entirely by the authors.




\bibliography{aaai25}
\clearpage

\end{document}